\documentclass[12pt,epsfig]{article}
\usepackage{amsmath}
\usepackage{graphicx}
\usepackage{epstopdf}
\usepackage{lscape}
\usepackage{longtable}
\usepackage{float}
\usepackage[utf8]{inputenc}
\usepackage[usenames, dvipsnames]{color}
\newtheorem{definition}{Definition}[section]
\begin{document}
	\begin{center}
		\textbf {A new completely parameter-free clustering algorithm for unsupervised classification of BATSE gamma-ray bursts}\\
	\end{center}
	\begin{center}
		Dr. Soumita Modak\footnote[1]{Email: soumita.stats@presiuniv.ac.in; soumitamodak2013@gmail.com\\
			Orcid id: 0000-0002-4919-143X\\
			Website: sites.google.com/view/soumitamodak}\\
		Department of Statistics, Presidency University\\
		86/1 College Street, Kolkata-700073, India\\
	\end{center}
	Abstract:  Cluster analysis is a widely applied machine learning technique to understand the existing patterns in the population of gamma-ray bursts (GRBs), in order to explore their physical sources. In the present scenario, the number of clusters corresponding to differentiable groups is still under conflict, in spite of numerous attempts with the state-of-the-art clustering procedures. This crucial unknown parameter needs to be evaluated, either directly or indirectly in terms of other tuning parameters, to produce the clusters in GRBs through implementation of an appropriate clustering algorithm. While most of the applied algorithms reached two physically explained groups of merger and collapsar predominated by the short and long bursts respectively, other statistical approaches violated this binary partition. However, physical establishment of any additional cluster(s) is not yet confirmed. Therefore, we propose a new algorithm, from a different stream of clustering referred to as `completely parameter-free',  which carries out the classification of GRBs in a manner that has not been tried so far. It indicates two main groups, of short and long duration bursts from the BATSE sample, compatible with the merger-collapsar theory.
	\section{Introduction}
	Classification is a statistical machine learning method to study the pattern in an observed data set, wherein we classify the data members into a number of groups, homogeneous within themselves and heterogeneous from each other. The most challenging scenario occurs when we do not have any kind of information on these existing classes, usually not even the number of them; then the corresponding classification is done in an unsupervised way, known as cluster analysis, where the resulting groups are termed as clusters accounting for the explicable factors behind this grouping (Anderberg 1973; Hartigan 1975; Ripley 1996; McLachlan and Peel 2000; Duda, Hart and Stork 2001; Everitt, Landau and Leese 2001; Kaufman and Rousseeuw 2005; Johnson and Wichern 2007; Modak 2019; Modak, Chattopadhyay \& Chattopadhyay 2022; Modak 2024a; Sabarish et al. 2025). 
	
	This article considers the clustering problem of GRBs, with an attempt to solve the conflict around the true number of clusters existing in the population, that can explain all different physical sources responsible for creation of bursts with varying characteristics. Till date, the majority of cluster analyses have revealed mainly two groups of short and long duration GRBs, with hard and soft bursts respectively indicated by their spectral hardness ratios, where the latter contains brighter ones possessing higher fluence and peak flux variables (Mazets et al. 1981;
	 Dezalay et al. 1992; Kouveliotou et al. 1993; \v{R}\'{\i}pa et al. 2012; Yang et al. 2016; Kulkarni and Desai 2017; Tarnopolski 2019; Tarnopolski 2022; Modak 2025). 
	 The physical processes behind them are explored in view of prompt-emission, afterglow, host galaxy, redshift distribution, or other observational properties like association with supernova, kilonova, and
	 gravitation wave (Gehrels et al. 2009; Zhang et al. 2012; Berger 2014; Levan et al. 2016; Goldstein et al. 2017; Wang et al. 2017; Lamb et al. 2019; Melandri et al. 2019; Troja et al. 2019; Jin et al. 2020; Minaev \& Pozanenko 2020; Zhu et al. 2024), which led to believing that short bursts are generally have compact binary merger and long bursts have massive stellar collapse as their progenitors (Paczy\'{n}ski 1986; Usov 1992; Woosley 1993; Paczy\'{n}ski 1998; Bloom et al. 2006; Woosley \& Bloom 2006; Nakar 2007; Berger 2014; Blanchard et al. 2016). For instance, an individual GRB associated with a supernova is known to originate from the core collapse of a massive star (Melandri et al. 2019; Minaev \& Pozanenko 2020), and that with a kilonova from the merger of binary neutron stars (Wang et al. 2017; Lamb et al. 2019); however, the correspondence of short bursts to merger and long to collapsar is violated by evidences of the reverse origins for some of the unusual bursts, but no distinct progenitor could thus far be explained for them (Della Valle et al. 2006; Zhang et al. 2009; Zhang et al. 2012; Ahumada et al. 2021; Rossi et al. 2022; Troja et al. 2022; Zhu et al. 2022; Levan et al. 2024; Yang et al. 2024; Wei et al. 2026).
	
	On the other hand, an additional intermediate group, by means of duration only or along with other variables, is suspected
	(Horv\'{a}th 1998, 2009; Mukherjee et al. 1998; Balastegui et al. 2001;  Horv\'{a}th et al. 2006, 2018; Tsutsui \& Shigeyama 2014; Horv\'{a}th \& T\'{o}th 2016; \v{R}\'{\i}pa \& M\'{e}sz\'{a}ros 2016; Tóth et al. 2019; Ghosh 2025), which might be linked to the merger of a massive white dwarf with a neutron
	star (Chattopadhyay et al. 2007; King et al. 2007; Modak et al. 2018; Yang et al. 2022). However, physical existence of this third group is not confirmed yet that could have arisen spuriously due to the instrumental and
	sampling biases or observational selection effects (Hakkila et al. 2000; 2003; Rajaniemi \& M\"{a}h\"{o}nen 2002;  Zitouni et al. 2015; Zhu et al. 2025; Wei et al. 2026). In contrast, more than three groups like four (Mehta and Iyyani 2024) or five (Chattopadhyay \& Maitra 2017, 2018; Acuner \& Ryde 2018) are proposed, whose possibility of possessing any separate progenitor is not discussed. Such outcome is increasing the confusion regarding the true number of meaningful groups (denoted by $K$) in GRB population, which hinders the search of different truly existing potential physical sources behind their occurrences. While (Toth et al. 2019) explored insignificance of such extra clusters arguing with their domain knowledge, Modak (2021) showed through fuzzy clustering that these further groups are not supported for existence. Then again, Modak (2025) prominently proved that the aforementioned misleading estimates of $K$ have occurred due to the implementation of inappropriate cluster accuracy measures like BIC or ICL (Frayley and Raftery 1998; Biernacki, Celeux and Govaert 2000), which determines the value of $K$ in the chosen finite-mixture-model based clustering algorithm; nevertheless, accurate measures, in association with the same clustering algorithm, lead to two familiar groups (see, Modak 2025, for details). It clearly demonstrates that $K$ is a crucial parameter in clustering of GRBs. Therefore, in this article, we propose a completely parameter-free clustering algorithm for unsupervised classification of GRBs, that automatically estimates the unknown $K$ without any intervention of the user, which results in two well-known clusters as well. 
	
	The rest of the paper is organized as follows. Section 2 outlines the existing streams of clustering methods. Section 3 describes our novel algorithm, under the paradigm of parameter-free methods, with its properties and applicability. Section 4 is dedicated to clustering of GRBs by the proposed method in superiority to the other earlier applied algorithms from the literature. We draw the conclusion in Section 5.
	\section{Clustering algorithm}
	We categorize the vast kinds of existing algorithms available for clustering into two broad categories: (i) these are common in the literature that depend on some tuning parameter(s), whereas (ii) the others are relatively newly emerged that do not depend on any sort of parameters and comparably much less in number than the former type. 
	\subsection{Parameter-dependent clustering algorithm}
	Cluster algorithms are implemented to cluster the data into $K$ (say) number of clusters where the true value of $K$ is generally unknown. Undoubtedly, the estimation of $K$ is crucial and requires special attention, while any of its incorrect estimates makes the entire classification misleading. So far, the analysts have been restricted mainly to the algorithms in need of specification for $K$ (i) either directly or indirectly, and (ii) prior, post or during the implementation of the algorithm (for details, see, Pakhiraa, Bandyopadhyay and Maulik 2004; Handl, Knowles and Kell 2005; Silva et al. 2020; Modak 2022; Modak 2024b; Modak 2024d). Let us illustrate this matter with some classical clustering approaches: (a) partitioning or model-based methods require $K$ to be provided to perform the clustering; and (b) hierarchical algorithms build a whole hierarchy of clustered data members, where we have singleton clusters to one group of all members constructed, from which a provided value for $K$ extracts the requested classification; while (c) density-based clustering determines the value of $K$ during the implementation of the algorithm itself and finally generates the partition. It is to be noted that the advanced clustering methods, that are intended to be improvements over the classical ones in the form of their combination or adaptation: like OPTICS, HDBSCAN, spectral clustering, clustering based on minimum spanning tree, neural network clustering and so on (see, Ankerst et al. 1999; Ng, Jordan and Weiss 2001; Herrero, Valencia and Dopazo 2005; Campello, Moulavi and Sander 2013; Lv et al. 2018; Shahid 2023; and references therein), also involve the aforementioned critical issue about assessment of the value of $K$. Therefore, all clustering processes need $K$ as a (tuning) parameter to be provided, which is performed either directly as in (a) and (b), or indirectly like in (c), as follows: (a) prior to the algorithm's implementation, (b) post-running the algorithm, and (c) during the process of the algorithm in terms of other involved parameter(s). To name a few popular classical instances: under (a) $K$-means or $K$-medoids method, Gaussian mixture model (GMM)-based clustering; (b) agglomerative or divisive hierarchical algorithms based on various linkages like average, single or complete; (c) DBSCAN or kernel-density based algorithms; and so on (Ester et al. 1996; McLachlan and Peel 2000; Kaufman and Rousseeuw 2005; Johnson and Wichern 2007; Matioli et al. 2018; Modak 2024c). 
	
	Evidently the two types of methods from (a) and (b) belong to the kind of clustering algorithms that needs $K$ as a priori, and therefore, necessitates the specification of $K$ directly; on the other hand, the methods in (c) fall into another kind of algorithms that does not require $K$ as a priori. The latter apparently seems to be advantageous over the former; however, involves other crucial tuning parameters that eventually determine the value of $K$. If the algorithms like $K$-means/medoids (Hartigan 1975; Kaufman and Rousseeuw 2005) ask for $K$ (a single direct parameter mention) as a prerequisite; DBSCAN (Ester et al. 1996), that groups the given sample into the algorithm-induced number of clusters, also requires $K$ to be known implicitly in the form of $K(minPts,\epsilon)$, i.e. as a function of two other important hyperparameters `$minPts$' and `$\epsilon$' both unknown and to be evaluated carefully. In DBSCAN algorithm, a minimum of $minPts$ data members is needed within the $\epsilon$ distance around a seed data member to form any cluster, where $K$'s value is returned at the termination of the algorithm for given values of its pair of arguments: $(minPts,\epsilon)$. Likewise, the clustering proposed in Matioli et al. (2018) or Modak (2023b) reveal $K$ at the conclusion of the process, but through implementation of the classical nonparametric kernel density estimation procedure that possesses an unknown bandwidth parameter $(h)$, whose value constructs the clusters and thereby eventually determines the number of clusters $K$. Any kind of change in the values of these other hyperparameter(s) may result in significantly different $K$ estimates. 
	
	Therefore, none of these algorithms facilitates the estimation of $K$ automatically without mediation by the analyst. Thus, when the authors refer to their methods as `not requiring $K$ as a priori or estimating $K$ during the clustering algorithm itself', it does not guarantee to be advantageous over the methods that necessitate $K$ being provided. Because we need to find out either of these parameters: $K$ or other(s) estimating $K$, by manual or automatic techniques (Ankerst et al. 1999; Koonsanit et al. 2012; Campello, Moulavi and Sander 2013; Hahsler et al. 2019; Mu et al. 2020; Modak 2023a) such as evaluating $K$ in $K$-means clustering (Koonsanit et al. 2012), or suggesting the pair $(minPts,\epsilon)$ in DBSCAN algorithm (Hahsler et al. 2019), and so on. Nevertheless, such approaches are all case-dependent, complex and experimental. 
	Moreover, they do not ensure the optimal results. Most importantly, they may produce unacceptable outcome with unreasonable clusters. Therefore, it is often recommended to select viable values of the parameters for clustering a given sample by suitable cluster validity indices or accuracy measures (Rousseeuw 1987; Frayley and Raftery 1998; Handl, Knowles and Kell 2005; Modak 2023b; Modak 2023c; Modak 2024b), that generally takes place in the following manner. Hierarchical clustering techniques (e.g. those mentioned in b), for distinct plausible values of $K$, obtain the respective clustering outcomes and finalize an $K$ as the estimate that corresponds to the optimum clustering results by means of cluster accuracy measure; whereas non-hierarchical ones (include those from a and c) involve the estimation of $K$ or other critical parameters to find $K$ from a credible range, that entails multiple occurrences of the entire algorithm over different possible choices of the concerned parameters to reach an optimum partition in terms of cluster validity index. In both the cases, the results depend heavily on the selection of the parameters and any wrong choice may lead to a misleading or absurd answer. These facts prominently explain the difficulty in parameter-selection, time-consuming multiple implementations of the algorithm, or choosing the appropriate way for validation of the resulting partition in the existing parameter-dependent clustering methods.
	\subsection{Parameter-free clustering algorithm}
	In an attempt to avoid the complicated procedure for determination of the unknown parameters engaged in the concerned clustering algorithm, cluster analysts are trying to establish a new paradigm of clustering methods that are completely parameter-free. For example, Chen et al. (2011) built a parameter-free approach by combining Affinity Propagation
	(AP) algorithm and an extended version of DBSCAN (i.e. DDBSCAN) to obtain arbitrary-shaped clusters with varying densities, but demonstration of its application was restricted to bivariate data; Rahman et al. (2018) proposed parameter-independent density-based clustering (PIDC) algorithms, however, their high-dimensional performance was not studied. In contrast, Gheyas et al. (2021) tried to propose an algorithm entirely parameter-free, although it depends on a user-specified parameter affecting the whole clustering, where the ratio of two adjacent clusters when found less than a threshold, which is a vital tuning parameter, merges the aforesaid clusters. This threshold is specified as two in their project without exploration of its variation with respect to different sets of data. There are many such works proposed as parameter-free, though they are actually not and find the concerned parameter values by automation techniques which can never entitle them to become parameter-free. For instance: Koonsanit et al. (2012) made an automated design for satellite imagery application to insert $K$ in the $K$-means algorithm, Mu et al. (2020) devised a new algorithm for selecting different density-level parameters in DBSCAN, Huang and Zeng (2025) developed a nearest neighbors-based clustering with an adaptive mechanism for tuning parameter $\gamma$, are eventually not parameter-free. Consequently, the authors' efforts to make these algorithms independent from  parameters have clearly failed.
	
	On the other hand, we discover a new clustering algorithm, neither requires $K$ nor any other parameters to be evaluated, and hence definitely qualifies to belong to the new paradigm of clustering methods entirely parameter-independent. Compared to the other stream of cluster algorithms, it does not need time-consuming multiple runs or exploration of resulting clusters for various considered values of the parameters to reach a best one among them. This method is certainly beneficial not involving any difficult parameter-selection procedure, which objectively divides the data in a step-wise process, where a suitable clustering accuracy measure assures the termination of the algorithm resulting in the algorithm-induced number of well-defined clusters with arbitrary sizes, shapes and densities.
	\section{Proposed clustering algorithm}
	\subsection{Some definitions}
	\begin{definition}[Partition]\label{def:partition}
		For a given data set (or sample) $D$ in $p$-dimensional real space with $n$ members: $D=\{M_i\in R^p,i=1(1)n\}$ wherein $M_i$ is the $i$-th member,
		$P^K_{\cup_{k=1}^{K} C_{k}}(D)$ is called a partition of $D$ into $K(>1)$ clusters $\{C_k,k=1(1)K\}$ such that 
		\begin{equation}\label{clusterproperty}
			C_k\neq \phi \hspace{.05in} \forall k,	C_k \cap C_{k'}=\phi\hspace{.05in} \forall k\neq k' \hspace{.05in}\text{and}\hspace{.05in}
			\cup_{k=1}^{K} C_{k}=D,
		\end{equation}
		where			
		\begin{equation}\label{clustersize}
			Card(C_k)=n_k \hspace{.05in}\text{with}\hspace{.05in}\sum_{k=1}^{K}n_k=n.
		\end{equation}
		`$Card$' stands for the cardinality of a set.
	\end{definition}
	\begin{definition}[Interpoint distance]\label{def:ID}
		The interpoint distance between any two members $M_i$ and $M_j$ from the data set $D$ is defined as a distance measure computed between them as follows
		\begin{equation*}
			d(M_i,M_j): D \times D \rightarrow [0,\infty)
		\end{equation*}
		satisfying\\
		(i)	$d(M_i,M_j)\geq0\hspace{.05in} \forall i,j$ \hspace{.05in}\text{(non-negativity)}\\
		(ii) $d(M_i,M_j)=0$ \text{iff} $i=j$ (identity of indiscernibles)\\
		(iii) $d(M_i,M_j)=d(M_j,M_i)\hspace{.05in}  \forall i,j$ (symmetry).\\
		The above measure becomes a strictly distance metric when fulfills additionally the following property:\\
		(iv) $d(M_i,M_j)\leq d(M_i,M_k)+d(M_k,M_j)\hspace{.05in}  \forall i,j,k$ (triangle
		inequality).\\	 
	\end{definition}
	\begin{definition}[Interpoint distance matrix]\label{def:IDM}
		For the given data set $D=\{M_i,i=1(1)n\}$ with $Card(D)=n$, an $n \times n$ interpoint distance matrix is defined through all interpoint distances computed for the entire sample as
		\begin{equation*}
			DistMatrix=(d(M_i,M_j))_{i,j=1(1)n}.
		\end{equation*} 
		
	\end{definition}
	\begin{definition}[$l$-th ordered member]\label{def:lOM}
		A member from the data set is called the $l$-th ordered member, denoted by $M_{(l)}$, whose aggregate distance from the sample, that is defined to be the sum of its distances from all members, is of $l$-th order, i.e.
		\begin{equation*}
			rank\Bigg(\sum_{j=1}^{n}d(M_i,M_j)\Bigg)=l\hspace{.05in} when\hspace{.05in} M_i=M_{(l)},
		\end{equation*}
		where `rank' is computed as the order of $\sum_{j=1}^{n}d(M_i,M_j)$ in the series\\  $\big\{\sum_{j=1}^{n}d(M_i,M_j),i=1(1)n\big\}$ when arranged in an increasing order of values.	
	\end{definition}
	\begin{definition}[$l$-th nearest member from a member]\label{def:lNM}
		The $l$-th nearest member from member $M_i$ is the one possessing the $l$-th order distance from $M_i$, which is denoted by $M_{(i)_l}$, satisfying
		\begin{equation*}
			rank\big(d(M_{i},M_j)\big)=l\hspace{.05in} with\hspace{.05in}M_j=M_{(i)_l},
		\end{equation*}
		among the series $\{d(M_{i},M_j),j=1(1)n\}$ considered in an increasing order of values.
	\end{definition}
	\begin{definition}[Average silhouette width]\label{def:ASW}
		For a given partition $P^K_{\cup_{k=1}^{K} C_{k}}(D)$ of the sample, as per Definition~\eqref{def:partition}, we define the average silhouette width (ASW) for the entire data set as
		\begin{equation}\label{eq:asw}
			\text{ASW}\Big\{P^K_{\cup_{k=1}^{K} C_{k}}(D)\Big\}=\frac{1}{n}\sum\limits_{i=1}^{n}s(M_i),
		\end{equation}
		where 
		\begin{equation*}
			s(M_{i})=\frac{b(M_{i})-a(M_{i})}{\max\big\{a(M_{i}),b(M_{i})\big\}}
		\end{equation*}
		is the silhouette width for the $i$-th member $M_i$ from the sample. Suppose the member $M_i$ is clustered in the $k$-th cluster $C_k$, then
		\begin{equation*}
			a(M_{i})={\sum\limits_{j(\neq i)=1:M_j\in C_{k}}^{n_{k}}d(M_{i},M_{j})}\big{/}{(n_{k}-1)}
		\end{equation*}
		\text{and}
		\begin{equation*}
			b(M_{i})=\underset{k'(\neq k)=1(1)K} {\min}\Bigg\{{\sum\limits_{j=1:M_j\in C_{k'}}^{n_{k'}}d(M_{i},M_{j})}/{n_{k'}}\Bigg\}.
		\end{equation*}
		Note the $ASW\in [-1,1]$ (see, Eq.~\ref{eq:asw}) with a higher value indicates a superior clustering of the sample $D$. Consequently, the partition $P^K_{\cup_{k=1}^{K} C_{k}}(D)$ is indicated to be better than another partition $P^{K'}_{\cup_{k=1}^{K'} C'_{k}}(D)$ if
		\begin{equation}
			ASW\Big\{P^K_{\cup_{k=1}^{K} C_{k}}(D)\Big\}>ASW\Big\{P^{K'}_{\cup_{k=1}^{K'} C'_{k}}(D)\Big\}\hspace{.05in}\text{for}\hspace{.05in}K=K'\hspace{.05in}\text{or}\hspace{.05in}K\neq K',
		\end{equation}
		(see, Rousseeuw 1987; Kaufman and Rousseeuw 2005; Modak 2019).	
	\end{definition}
	\subsection{Algorithm}\label{Algo}
	Objective: Assume the given sample has some inherent clustering structure with unknown $K(\geq2)$ number of clusters. We develop the following algorithm in a parameter-free approach, consisting of two phases, which determines $K$ and results in $K$ well-defined clusters. \\
	
	\textbf{Phase 1}:
	\begin{enumerate}
		\item We have the given data set: 
		\begin{equation}\label{data}
			D=\{M_i,i=1(1)n\}\hspace{.05in}\text{with}\hspace{.05in}Card(D)=n.
		\end{equation} 
	\item For $D$, calculate the $n\times n$ interpoint distance matrix as per Definition~\eqref{def:IDM},
		where `$d$' is a user-chosen interpoint distance measure (see, Definition~\ref{def:ID}).
		\item Determine the first ordered data member from $D$ by Definition~\eqref{def:lOM}: 
		\begin{equation*}
			M_{(1)}=\underset{i=1(1)n}{\arg \min}\sum_{j=1}^{n}d(M_{i},M_j)
		\end{equation*}
		or equivalently
		\begin{equation*}
			M_{(1)}=\underset{i=1(1)n}{\arg\min}{\sum_{j(\neq i)=1}^{n}}d(M_{i},M_j), 
		\end{equation*}
		which is likely to be the most centrally situated data member among all in the sample.  
		\item Find the $l$-th nearest member from the member $M_{(1)}$, by Definition~\eqref{def:lNM}, for $l=1(1)n$: 
		\begin{equation*}
			\{M_{(1)_l},l=1(1)n\} \ni 			d(M_{(1)},M_{(1)_{l'}})<d(M_{(1)},M_{(1)_{l''}})\hspace{.05in}\forall\hspace{.05in} l'<l'',
		\end{equation*}
		where 
		\begin{equation*}
			M_{(1)_l}=M_{(1)}\hspace{.05in}\text{for}\hspace{.05in}l=1,
		\end{equation*}
		and $M_{(1)_n}$ is the furthest member of the sample from the centrally located one $M_{(1)}$.
		\item Construct the first two (initial) clusters as follows: 
		\begin{equation}\label{C1data}
			C'_1=\{M_{1_j},j=1(1)n'\} \hspace{.05in}\text{with}\hspace{.05in}Card(C'_1)=n'
		\end{equation}
		and
		\begin{equation}\label{C2data}
			C'_2=\{M_{2_j},j=1(1)n-n'\} \hspace{.05in}\text{with}\hspace{.05in}Card(C'_2)=n-n' 
		\end{equation}
		such that the following two conditions, namely C1 and C2, are satisfied.
		\begin{equation}\label{C1}
			\begin{aligned}
				\hspace{.05in}\text{(C1):}\hspace{.1in}d(M_{(1)_1},M_{1_j})<d(M_{(1)_n},M_{1_j})\hspace{.05in}\text{for}\hspace{.05in}j=1(1)n'
			\end{aligned}	 	
		\end{equation}
		and 
		\begin{equation}\label{C2}
			\begin{aligned}
				\hspace{.05in}\text{(C2):}\hspace{.1in}d(M_{(1)_1},M_{2_j})>d(M_{(1)_n},M_{2_j})\hspace{.05in}\text{for}\hspace{.05in}j=1(1)n-n'.
			\end{aligned} 	
		\end{equation}
		That is, by condition C1 from Eq.~\eqref{C1}, we build our first cluster $C'_1$ with $(n'-1)$ data members clustered around the member $M_{(1)_1}$, and condition C2 in Eq.~\eqref{C2} leads to the second cluster $C'_2$ of size $(n-n')$ having its center close to the member $M_{(1)_n}$, so that the pair of clusters are far from each other. 
		
		\item The above step produces a partition of $D$ into two clusters: $P^2_{C'_1\cup C'_2}(D)$ whose quality is assessed by the $ASW$ from Definition~\eqref{def:ASW}: 
		\begin{equation}\label{ASW1}
			ASW\{P^2_{C'_1\cup C'_2}(D)\}.
		\end{equation}
		\item Next, we investigate if the formed clusters can be further clustered to achieve a better partition with respect to Eq.~\eqref{ASW1} as follows.
		
		In step (1), instead of the entire sample from Eq.~\eqref{data}, we now start with the members from $C'_1$ as in Eq.~\eqref{C1data}, and repeat steps (1-5) to find two sub-clusters out of the initial cluster $C'_1$, namely $C'_{11}$ and $C'_{12}$. Then if 
		\begin{equation}
			ASW\{P^2_{C'_1\cup C'_2}(D)\}\geq ASW\{P^3_{C'_{11}\cup C'_{12}\cup C'_2}(D)
			\},
		\end{equation}
		the clustering is not considered to be improved by this division; therefore, we discard the obtained partition and retain the original cluster $C'_1$ as it is. Otherwise, we accept the grouping of $C'_1$ into two, and proceed with looking for further sub-groups in each of the resulting clusters. This break-up process continues till no more improvement in the clustering is suggested by means of the $ASW$.
		
		Suppose we finally reach a partition of $n'$ data members into $K_1$ clusters resulted from the initial cluster $C'_1$ as follows: 
		\begin{equation}\label{C1partition}
			P^{K_1}_{\cup_{k=1}^{K_1} \tilde{C}_k}(C'_1)= \{\tilde{C}_k,k=1(1)K_1\},
		\end{equation}
		where
		\begin{equation}
			ASW\Big\{P^{K_1+1}_{\cup_{k=1}^{K_1} \tilde{C}_k\cup C'_2}(D)\Big\}
		\end{equation} 
		can not better by additional classification of the partition from Eq.~\eqref{C1partition}.
		\item Redo the preceding step (7) with the other initial cluster $C'_2$, having members from Eq.~\eqref{C2data}, in an attempt to obtain another branch of sub-clusters out of  $C'_2$.
		
		Let we eventually arrive at a partition of $(n-n')$ data members into $K_2$ resulting clusters out of the initial cluster $C'_2$ as follows: 
		\begin{equation}\label{C2partition}
			P^{K_2}_{\cup_{k=K_1+1}^{K_1+K_2} \tilde{C}_k}(C'_2)= \{\tilde{C}_k,k=K_1+1(1)K_1+K_2\},
		\end{equation}
		where
		\begin{equation}
			ASW\Big\{P^{K_2+1}_{C'_1\cup_{k=K_1+1}^{K_1+K_2} \tilde{C}_k}(D)\Big\}
		\end{equation} 
		can not be improved by further division of the partition in Eq.~\eqref{C2partition}.
	\end{enumerate}	
	\textbf{Phase 2}:\\ 
	
	Next, we reassign the clustered members obtained from phase (1), if needed, so that they all fall in a cluster with the closet mean, which is computed with respect to the clustering of members as resulted in phase (1).
	
	Phase (1) results in the following partition of $D$ into $K_1+K_2=K$ clusters:
	\begin{equation}
		P^K_{\cup_{k=1}^{K} \tilde{C}_k}(D)=\{\tilde{C}_k,k=1(1)K\},
	\end{equation}
	for which we define the cluster-wise mean as the average of all members from each cluster $\tilde{C}_k$ given below:
	\begin{equation}
		Mean_{\tilde{C}_k}=\frac{1}{Card(\tilde{C}_k)}\sum_{i=1:M_i\in\tilde{C_k}}^{Card(\tilde{C}_k)}M_i\hspace{.05in}\text{for} \hspace{.05in}k=1(1)K.
	\end{equation}
	Then, the distance between any data member of the sample $M_i$ and a resulting cluster $\tilde{C}_k$ is defined by
	\begin{equation}
		d(M_i,\tilde{C}_k)=d(M_i,Mean_{\tilde{C}_k})\hspace{.05in}\text{for} \hspace{.05in}i=1(1)n,k=1(1)K.
	\end{equation}
	At the end of phase (2), we make sure that any $M_i$ is assigned to a cluster (say $C_{k'}$) if and only if
	\begin{equation}\label{min}
		C_{k'}=\underset{k=1(1)K} {\arg\min}\hspace{.05in}d(M_i,\tilde{C}_k),
	\end{equation}
	whereas if an $M_i$ has already achieved the property defined by Eq.~\eqref{min} in phase (1), then no reshuffle for it is needed in phase (2). 
	
	Thus, we reach our final partition of the given sample with well-constructed clusters:
	\begin{equation}\label{finalpartition}
		P^K_{\cup_{k=1}^{K} C_k}(D)=\{C_k,k=1(1)K\} \hspace{.05in}\text{with} \hspace{.05in}Card(C_k)=n_k\ni \sum_{k=1}^{K}n_k=n,
	\end{equation}
	where $C_k=\tilde{C}_k \forall k$ if no relocation takes place in phase (2), otherwise $C_k\neq\tilde{C}_k$ for at least two $k$ s.
	
	Quality of the clustering attained through our algorithm in Eq.~\eqref{finalpartition} is immediately evaluated by
	\begin{equation}
		\text{ASW}\Big\{P^K_{\cup_{k=1}^{K} C_k}(D)\Big\}\in[-1,1],
	\end{equation} 
	whose higher value is considered as a manifestation of a better classification of the data that can be compared with any other partition of this sample obtained from a different clustering algorithm.
	\subsection{Properties of the algorithm}
	\begin{enumerate}
		\item Our algorithm is independent of the order of insertion of members in the data set (see, step 1, Section~\ref{Algo}).
		\item The algorithm initiates itself objectively without the user's subjective intervention: that avoids multiple random initializations, required in algorithms such as $K$-means or Clara (Hartigan 1975; Kaufman and Rousseeuw 2005), or any extensive procedure to find the initial clusters, like PAM while implementing $K$-medoids method (Modak, Chattopadhyay \& Chattopadhyay 2017; Modak, Chattopadhyay \& Chattopadhyay 2020).
		\item This is completely parameter-free neither needs $K$ nor any other parameters to be provided unlike most of the algorithms, classical or adaptive, existing in the literature.
		\item It constitutes of two phases: the first determines $K$ clusters, and then the second performs any possible improvements. Hence, at the end we achieve $K$ well-defined clusters that do not change with the slightest shift in the values of parameters engaged in other established clustering algorithms. For example, methods such as $K$-means, Clara or spectral clustering, even for a fixed $K$, produce different partitions during their individual implementations due to the randomization involved in the algorithms (Hartigan 1975; Jordan and Weiss 2001; Kaufman and Rousseeuw 2005); whereas the density-based methods also suffer from this concern for varying values of their participating parameter(s) other than $K$ (Ester et al. 1996; Matioli et al. 2018; Modak 2023a); or model-based clustering facing the same issue for a given $K$, with extensive estimation of the concerning model-parameters, that involves iterative approach over multiple restarts to find a good solution (McLachlan and Peel 2000; Surucca et al. 2016).
		\item Our approach is capable of identifying clusters with varying densities. It is a very vital quality of a clustering method that many popular algorithms lack; in fact, the efficient density-based methods like DBSCAN do not meet this quality due to its global parameter duo $(minPts,\epsilon)$ that are fixed over all clusters (Ester et al. 1996), whereas algorithms like DENCLUE, in spite of using local density maxima, fail because of the same choice of value for the involving hyperparameters in every cluster (Hinneburg and Keim 2003). In contrast, our algorithm, from its very initial phase to the end, takes care of the aforesaid concern through neighborhood-based concept by studying the order of proximity among the members in terms of their interpoint distances, based on which we implement the theory of nearest neighbor. Moreover, being parameter-free, our approach works without fixing any such parameters for the clusters that can restrain it from exposing a partition with groups of arbitrary densities.
		\item Often clustering methods are intended to construct uniformly shaped clusters, for instance: $K$-means or $K$-medoids algorithms find groups of spherical shapes, and GMM-based clustering targets some specific-shaped clusters according to the structure of dispersion matrices for contributing component densities, whereas hierarchical algorithms provide the clusters in the design of a tree aimed at extracting particular-shaped groups depending on an individual linkage method used (Kaufman and Rousseeuw 2005; Johnson and Wichern 2007; Surucca et al. 2016). However, our approach facilitates arbitrary-shaped clusters. 
		\item The later phase of our algorithm gives a chance of improvement over the already built clusters from the former phase through reshuffling, which is not feasible in algorithms similar to hierarchical ones, that is why they are heavily sensitive to noise, outlier or extreme observation.
		\item Algorithm always converges in contrast to $K$-means method (Hartigan 1975) which does not guarantee the convergence with arbitrary initialization of the cluster-centroids, or kernel-based algorithms: for example, ClusterKDE (Matioli et al. 2018) that fails to converge for an inappropriate value of the bandwidth parameter involved in the implemented kernel density estimation method (Bandyopadhyay and Modak 2018), or non-classical adaptive algorithms like spectral clustering (Jordan and Weiss 2001), and so on.
		\item This interpoint distance-based algorithm is applicable to all kind of data measured on arbitrary scales by means of any appropriate measure of distance, dissimilar to $K$-means or finite mixture model-based clustering which are limited to specific kind of data, for instance: GMM-based clustering is usable for continuous data only (McLachlan and Peel 2000; Modak, Chattopadhyay \& Chattopadhyay 2018; Tóth, Rácz \& Horváth 2019).
		\item It is useful for any-dimensional data set, even in high-dimensional scenario where the cluster sizes are close to or less than the dimensionality of the data space. As our method involves only interpoint distance matrix and not the multivariate parameters like dispersion matrix or so, which are poorly fitted by the smaller number of observations from higher-dimensional space, that occurs in any finite mixture model-based clustering algorithms (Bai and Saranadasa 1996; Yata and Aoshima 2010; Marozzi 2015; Modak and Bandyopadhyay 2019).

	\end{enumerate}
		\subsection{Applicability of proposed method}
	This section demonstrates the applicability of our method in terms of two case-studies: one is the benchmark synthetic data set for clustering, and the other is well-known challenging real-life astronomical sample, under CS1 and CS2 respectively.
	
\begin{figure}
		\centering
		\includegraphics[width=1\textwidth]{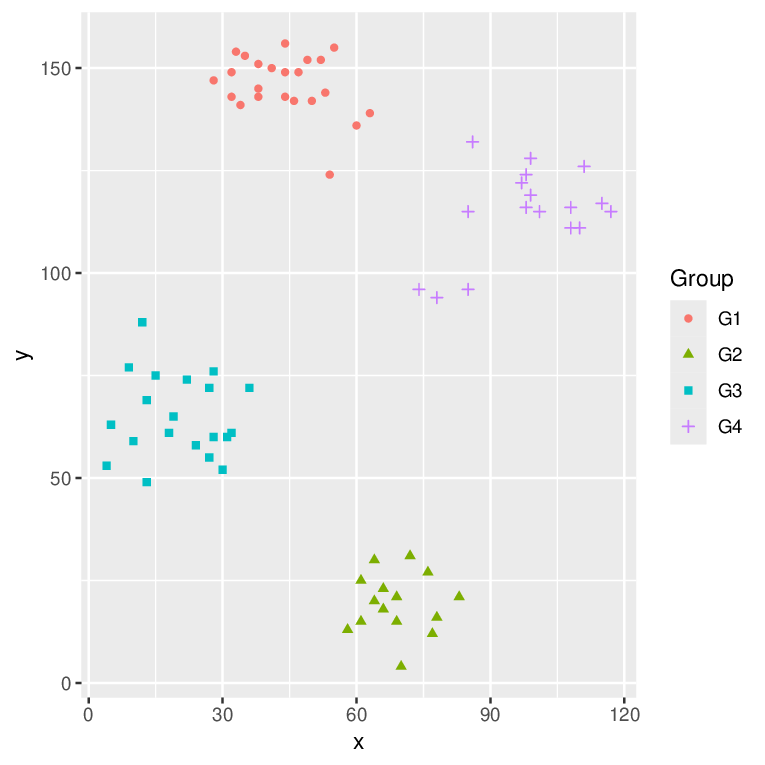}
		\caption{Ruspini data set with four arbitrary-shaped groups G1-G4.}\label{Ruspini}
	\end{figure}

	(CS1) Firstly, the effectiveness of our novel algorithm is explored through data set `Ruspini' (Ruspini 1970), which is a bivariate synthetic sample of size 75 with four arbitrary-shaped groups. 
	Our method exposes them with 100\% accuracy (see, Fig.~\ref{Ruspini}).
	
\begin{figure}
		\centering
		\includegraphics[width=1\textwidth]{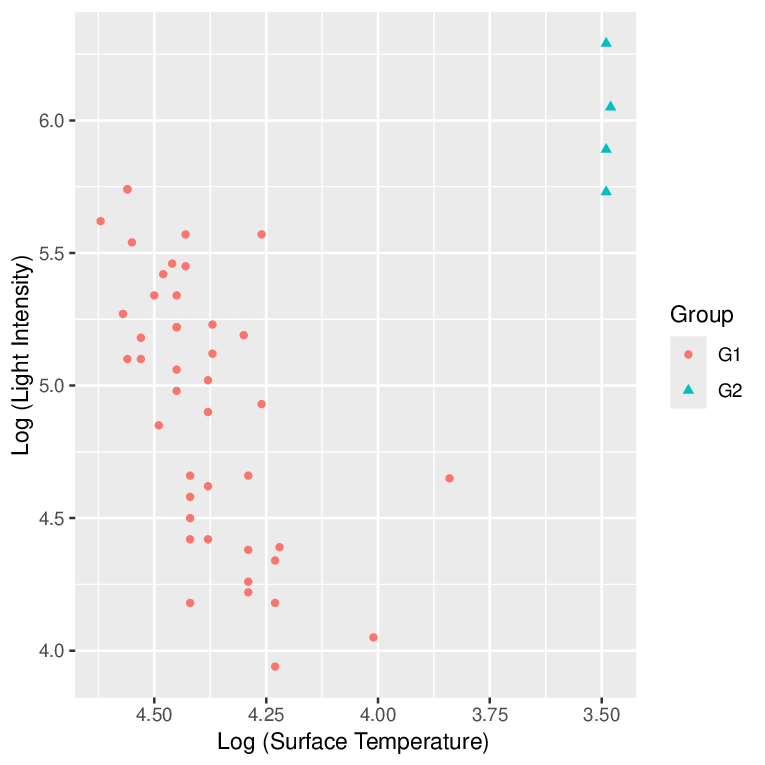}
		\caption{Scatter plot of the `CYG OB1’ star cluster data with two groups G1 and G2 of different start types.}\label{CYG-OB1}
	\end{figure}

	(CS2) Next is the sample from the star cluster `CYG OB1' possessing two groups: 43 stars from
	the main sequence and the other group is of 4 giant stars (Kaufman and Rousseeuw 2005). Fig.~\ref{CYG-OB1} shows the plot of the stars for study variables: surface temperature versus light intensity. The large difference in group sizes and its challenging clustering structure made most of the established algorithms unsuccessful in exposing the underlying grouping.  However, our proposed algorithm comes effective in retrieving the groups accurately with 100\% precision.
\section{GRB clustering}
In the present section, we apply our novel parameter-free clustering algorithm to GRB data set. The prescribed approach is ideal in this scenario as it itself solves the conflict around the value of $K$, in contrast to other popular methods implemented in the past that caused ambiguity regarding $K$ due to the evaluation procedures for the participating parameter(s). We also compare our method with some of those to prove its superiority in order to establish its acceptance with a higher degree. 
\subsection{Data set}
	We perform our analysis using the current data from the BATSE GRB Catalog\footnote{https://gammaray.nsstc.nasa.gov/batse/grb/catalog/current/} based on the following six variables:
	\begin{equation}\label{WV}
(log_{10}T_{50}, log_{10}T_{90}, log_{10}P_{256}, log_{10}F_{T}, log_{10}H_{32},log_{10}H_{321})
	\end{equation}
with finite values on 1956 bursts (Mukherjee et al. 1998; Hakkila et al. 2000; Chattopadhyay et al. 2007, Chattopadhyay and Maitra 2017; Toth et al. 2019; Modak 2021; Modak 2025), where the total fluence is given by 
\begin{equation*}
F_T=F_1+F_2+F_3+F_4,
\end{equation*}
and the spectral hardness ratios considered are 
\begin{equation*}
H_{32}=F_3/F_2\hspace{.05in}\text{and}\hspace{.05in} H_{321}=F_3/(F_2+F_1).
\end{equation*} 
$F_1,F_2,F_3,F_4$ represent time-integrated fluences respectively in $20-50$, $50-100$, $100-300$ and $>300$ keV spectral channels measured in ergs cm$^{-2}$; $P_{256}$ indicates the peak flux measured in $256$ ms bins with counts cm$^{-2}$ s$^{-1}$; $T_{50}$ and $T_{90}$ measure the duration in seconds (s) by which $50\%$ and $90\%$ of the flux arrive respectively.
 \subsection{Clustering algorithms}
 A total of three different clustering algorithms are implemented to the above data, where our parameter-free method is run without any parameter input or user's intervention that generates two clusters automatically. 
 
Secondly, we consider the $K$-means clustering method (Hartigan 1975), executed through the Hartigan-Wong algorithm (Hartigan and Wong 1979), that has been previously used to achieve the desired clustering pattern out of the GRBs (Chattopadhyay et al. 2007; Modak, Chattopadhyay \& Chattopadhyay 2018; Modak 2019; Modak 2025). Regarding the crucial parameters involved in the algorithm, namely the number of clusters and the initial cluster-centroids, we put $K=2$ and rerun the program with 100 random initial choices of the cluster-centroids to reach a stable answer.

Next, we consider the GMM-based clustering (McLachlan and Peel 2000), popularly employed to cluster GRBs (Mukherjee et al. 1998; Chattopadhyay \& Maitra 2017; Tóth et al. 2019; Modak 2025), where the most difficult parameter $K$ is provided to be two as a priori, and the other model parameters (in particular model VVV; see, Modak 2025) are determined by their maximum likelihood estimates through the EM algorithm (Scrucca et al. 2016).
 
\subsection{Comparative study}
We compare the partition of the bursts resulted in our algorithm with those from the aforesaid two techniques implemented earlier in the literature for the purpose of clustering of GRBs, and to achieve so we make use of the effective internal cluster validity index the `connectivity' (Handl et al. 2005), which is described below.

Suppose a random member from the sample $M_i$ is assigned to the cluster $C_k$, then we formulate:
$$
I_{i}(l)=\begin{cases}
	0 & \text{if $M_{(i)l}\in C_k$}, \\
1/l & \text{otherwise}.
\end{cases}
$$ 
The concerned measure connectivity is define by
\begin{equation}
	\text{Conn}_l=\sum\limits_{i=1}^{n}\sum\limits_{l=1}^{L}I_i(l), 
\end{equation}
which is a function of the parameter $l$. 
\begin{equation*}
0\leq\text{Conn}_l<\infty\hspace{.05in} \forall\hspace{.05in}  l=1(1)\underset{k=1(1)K}{\min}\{n_k-1\},
\end{equation*}  
where a larger value suggests a superior partition. 

Table~\ref{t1} reports this measure, for clustering output from three different algorithms, with varying values of $l$ to check the robustness with respect to this parameter; which shows consistent superiority for our advised methodology. Therefore, we explore the interpretation of these resulting groups attained through the proposed parameter-free clustering with respect to their study variables in the following section. 
\subsection{Results and discussions}
\begin{figure}
		\centering
		\includegraphics[width=1\textwidth]{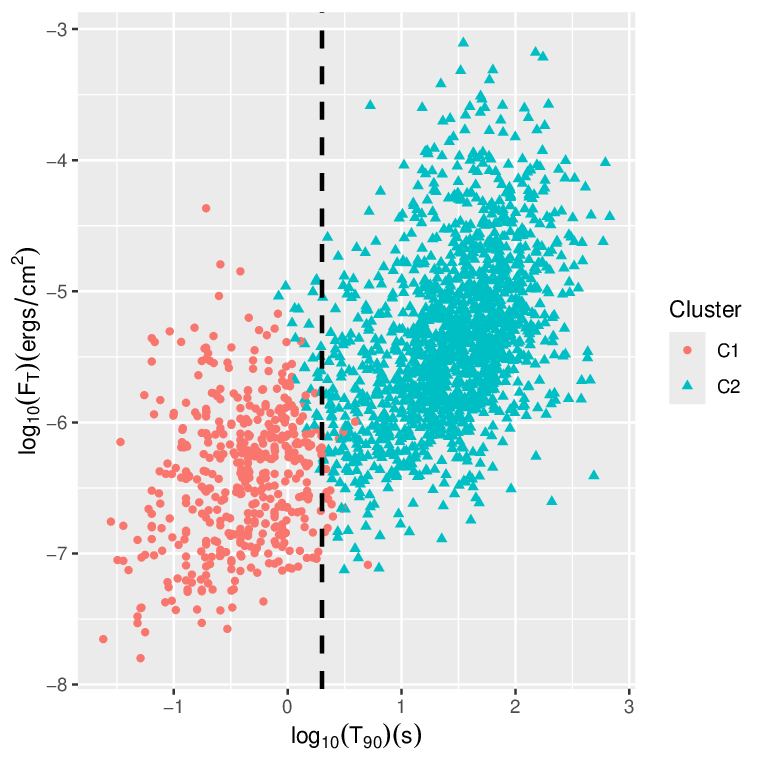}
		\caption{$log_{10}(T_{90})$ (in s) vs. $log_{10}(F_{T})$ (in ergs cm$^{-2}$) plot for two clusters of GRBs from the proposed parameter-free algorithm, wherein the vertical black dotted line represents $T_{90}=2$ s.}\label{TimevsFluence}
	\end{figure}

\begin{figure}
		\centering
		\includegraphics[width=1\textwidth]{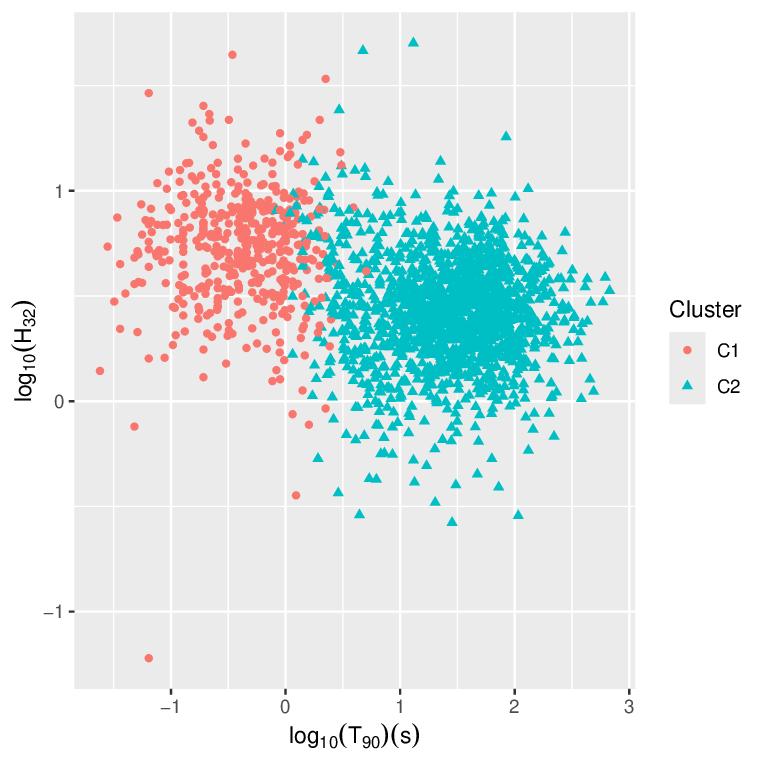}
		\caption{Plot of $log_{10}(T_{90})$ (in s) vs. $log_{10}(H_{32})$ for two clusters of GRBs derived from the proposed parameter-free algorithm.}\label{TimevsH}
	\end{figure}

Our proposed algorithm applied to the variables from Equation~\eqref{WV} produces two clusters of GRBs from the BATSE catalog. Table~\ref{t2} reports their average properties, and Figure~\ref{TimevsFluence} and Figure~\ref{TimevsH} respectively exhibit time versus fluence and time versus hardness ratio plots for individual bursts from both the clusters. 
The resulting clusters, namely C1 and C2, are prominently dominated by the short-hard-faint and long-soft-bright bursts respectively, with respect to their study variables in our algorithm that measure duration, hardness and brightness of the GRBs. These binary grouping is likely to be compatible with the merger-collapsar theory; however, further astrophysical analysis needs to be performed for the confirmation. 

Most of the empirical cluster analyses has led to the aforementioned cluster-duo predominated by the typical short and long duration bursts (Mazets et al. 1981;
Kouveliotou et al. 1993; Kulkarni and Desai 2017; Modak 2025), where the group with shorter bursts corresponds to compact binary mergers like merger of two neutron stars or merger of a neutron star with a black hole, and the longer ones respond to massive collapsars (Woosley \& Bloom 2006; Nakar 2007; Berger 2014; Blanchard et al. 2016). While an individual GRB's association with kilonova or supernova respectively confirm the progenitor as compact binary merger or core collapse of a massive star (Goldstein 2017; Melandri et al. 2019; Troja et al. 2019; Minaev \& Pozanenko 2020; Zhu et al. 2022; Levan at al. 2024); there are outliers or unusual bursts with hybrid traits violating one-to-one relation between short-long bursts and merger-collapsar modeling, however, no other progenitor is found yet for them (Bromberg et al. 2013; Zitouni et al. 2015; Barnes and Metzger 2023; Luo et al. 2023; Yang et al. 2024; Wei et al. 2026). We are well aware that there is significant overlap between the two groups predominated by short and long duration bursts in view of their multiple observational characteristics (Kouveliotou et al. 1993; Chattopadhyay et al. 2007; Modak, Chattopadhyay and Chattopadhyay 2018; Modak 2021), where the study variables are highly affected by the measuring instrument, energy band or other selection effects (Qin et al. 2013; Minaev \& Pozanenko 2020; Zhu et al. 2024). Still neither duration-free clustering (Wei et al. 2026), nor grouping based on the intrinsic properties taken into account the redshift of the bursts (Minaev \& Pozanenko 2020; Zhu et al. 2025), could justify any new progenitors. Cluster analysis of GRBs to find meaningful groups is only successful when it connects the clusters with possible distinct progenitors, which is known to be two till date, as indicated by most of the machine learning techniques applied so far including our novel parameter-free approach that automatically indicates two clusters.  
 \section{Conclusion}
 In the present paper, we consider a new paradigm of clustering algorithms that does not require specification of any kind of parameters. These methods are ideal for clustering a sample with the number of clusters unknown, that motivated us to develop a new such algorithm to classify GRBs in an unsupervised way without a single tuning parameter. Nevertheless, our suggested method is applicable to any astronomical data set for clustering purpose when parameters, like the number of existing groups or any other contributing to the parameter-dependent algorithms, are not only unknown but also difficult to estimate, wherein often even to guess at a plausible range for their values becomes impossible. For example, if we start searching for the unknown true value of $K$ in an interval that actually does not contain the original $K$, then we are in no position to reach the accurate scenario (Modak 2024d); whereas other parameters such as the bandwidth in a kernel-based clustering algorithm can possess unrealistic outcome as a manifestation of some of its unreasonably selected value (Matoli et al. 2018). To overcome similar hurdles, where no proper solution is available in the literature so far, parameter-free algorithm could be a good alternative. In this context, the present work proposes a new entirely parameter-free algorithm which is trained to produce automatically determined number of clusters without any intervention of the user by the use of the cluster accuracy measure ASW. Therefore, analysts are encouraged to explore the efficiency of our advised procedure incorporating other accuracy measures, provided the chosen one itself be free of any individual parameter(s) unlike nearest neighbor classification error rate (Ripley 1996), ICL (Biernacki, Celeux and Govaert 2000), connectivity (Handl et al. 2005), R$_{clus}$ (Modak 2022), and others.  
   
 Although we apply our method to GRB data set, it is in general useful for clustering arbitrary astronomical sample while addressing the issues arise in every other parameter-dependent algorithm. We demonstrate the proposed method's successful applicability through a synthetic data set, and one real-life challenging sample from astronomy as well. Nevertheless, our study on GRB in this article is performed using statistical modeling. Therefore, progenitors behind the resulting groups require to be confirmed with respect to further investigation in terms of their physical modeling and exploration of their cosmological properties. Moreover, the robustness of grouping structure in the GRB population should be checked in view of more data or samples from different catalogs than BATSE. We conclude with saying that a new clustering approach from parameter-free domain of algorithms is developed as a strong and efficient machine learning technique to cluster GRBs that indicated two groups, statistically better than found by earlier used methods such as $K$-means or GMM-based clustering (Modak 2025), which supports binary progenitor theory for the bursts' origination.
 
		\clearpage
		\begin{table}
			\caption{Comparative study of three algorithms (namely, the proposed, $K$-means and GMM-based) for clustering GRBs}
			\begin{center}
				\begin{tabular}{cccc}
					\hline
					Algorithm&$\text{Conn}_{10}$&$\text{Conn}_{20}$&$\text{Conn}_{30}$\\
					\hline
					Proposed&95.537&126.19&148.127\\
					$K$-means&100.313&136.54&161.013\\
					GMM-based& 149.291& 197.799& 231.593\\
					
					\hline
				\end{tabular}
			\end{center}
			\label{t1}
		\end{table}
		\clearpage
	\begin{table}
		\caption{Properties of two clusters (i.e. mean value with standard error) for GRBs resulted in our proposed parameter-free algorithm}
		\begin{center}
			\tiny
			\begin{tabular}{c c c c c c c c}
				\hline\\
				Cluster&Cluster-size & $T_{50}$ &  $T_{90}$ & $P_{256}$ & $F_{T} \times 10^{6}$ & $H_{32}$ &$H_{321} $\\
				name &  (percentage)        & (s)     & (s)     & (cm$^{-2}$ s$^{-1}$) &  (ergs cm$^{-2}$)      & &\\[1ex]
				\hline\\
				C1       & 452 (23.108\%)  & 0.233 $\pm$ 0.010& 0.626 $\pm$ 0.028& 2.165 $\pm$ 0.157& 0.909 $\pm$ 0.115&
				6.417 $\pm$ 0.213& 3.976 $\pm$ 0.110	\\[1ex]
				
				C2       & 1504 (76.892\%)  &   21.259 $\pm$ 1.018& 49.857 $\pm$ 1.672&  3.565 $\pm$ 0.223&  16.227 $\pm$ 1.262&  3.162 $\pm$ 0.066&  1.799 $\pm$ 0.033
				\\[1ex]
				\hline
			\end{tabular}
		\end{center}
		\label{t2}
	\end{table}
\clearpage
\clearpage
	\clearpage
	\clearpage
	

\clearpage
	\begin{thebibliography}{}
		\bibitem{}
		Acuner, Z. and Ryde, F. (2018). \textsl{Clustering of gamma-ray burst types in the Fermi GBM catalogue:
			indications of photosphere and synchrotron emissions during
			the prompt phase}. Monthly Notices of the Royal Astronomical Society. \textbf{475}, 1708--1724.
		\bibitem{}
		Aggarwal, C. C. and Reddy, C. K. (2014). \textsl{Data Clustering: Algorithms and Applications}.
		Chapman and Hall/CRC, New York. 
		\bibitem{}
		Ahumada, T., Singer, L. P., Anand, S. et al. (2021). \textsl{Discovery and confirmation of the shortest gamma-ray burst from a collapsar}. Nature Astronomy, \textbf{5}, 917–927 
		\bibitem{}
		Anderberg, M. R. (1973). \textsl{Cluster Analysis for Applications}. Academic Press, New York.
		\bibitem{}
		Anderson, T. W. (2003). \textsl{An Introduction to 
			Multivariate Statistical Analysis}, John Wiley \& Sons, Inc. New Jersey.
		\bibitem{}
		Ankerst, M., Breunig, M., Kriegel, H.-P. and Sander, J. (1999). \textsl{OPTICS: ordering points
			to identify clustering structure}. In Delis, A. et al. (eds), Proceedings of the 1999
		International Conference on Management of Data. ACM Press, New York, pp. 49–60.
		\bibitem{}
		Bai, Z., and H. Saranadasa (1996). \textsl{Effect of high dimension: By an example of a two sample problem}.
		Statistica Sinica, \textbf{6}, 311–329.
		\bibitem{}
		Balastegui, A., Ruiz-Lapuente, P., and Canal, R. (2001). \textsl{Reclassification of gamma-ray bursts.} Monthly Notices of the Royal Astronomical Society. \textbf{328}, 283--290.
		\bibitem{}
		Bandyopadhyay, U. and Modak, S. (2018). \textsl{Bivariate density estimation using normal-gamma kernel with application to astronomy}. Journal of Applied Probability and Statistics, \textbf{13}, 23-39.
		\bibitem{}
		Barnes, J., and Metzger, B. D. (2023). The Astrophysical Journal. \textbf{947}, 55.
		\bibitem{}	
		Biernacki, C., Celeux, G., and Govaert, G. (2000). \textsl{Assessing a mixture model for clustering with the integrated completed likelihood}. IEEE Trans. Pattern Analysis and Machine Intelligence, \textbf{22}, 719-725.
		\bibitem{}
		Blanchard, P. K., Berger, E., Fong, W.--f. (2016). \textsl{The Offset and Host Light Distributions of Long Gamma-Ray Bursts: A New View from HST Observations of Swift Bursts}. The Astrophysical Journal. \textbf{817}, 144.
		\bibitem{}
		Bloom, J. S., Prochaska, J. X., Pooley, D., Blake, C. H., Foley, R. J., Jha, S., Ramirez-Ruiz, E., Granot, J., Filippenko, A. V., Sigurdsson, S., Barth, A. J.,   Chen, H.-W., Cooper, M. C.,
		Falco, E. E., Gal, R. R., Gerke, B. F., Gladders,  M. D.,  Greene, J. E., Hennanwi, J., Ho, L. C., Hurley, K., Koester, B. P., Li, W., Lubin,  L., Newman,  J.,
		Perley, D. A., Squires, G. K. \& Wood-Vasey, W. M. (2006). \textsl{Closing in on a short-hard burst progenitor: constraints from early-type optical imaging and spectroscopy of a possible host galaxy of GRB 050509b}. The Astrophysical Journal. \textbf{638}, 354–-368.
		\bibitem{}
		Berger, E. (2011). \textsl{The environments of short-duration gamma-ray bursts and implications for their progenitors.} New Astronomy Reviews. \textbf{55}, 1--22.
		\bibitem{}
		Berger, E. (2014). \textsl{Short-Duration Gamma-Ray Bursts.} Annual Review of Astronomy and Astrophysics. \textbf{52}, 43--105.
		 \bibitem{}
		Bromberg, O., Nakar, E., Piran, T., \& Sari, R. (2013). \textsl{Short versus long and  collapsars versus non--collapsars:
			a quantitative classification of gamma-ray bursts.} The Astrophysical Journal. \textbf{764}, 179.
		\bibitem{}
		Campello, R. J. G. B., Moulavi, D., and Sander, J. (2013). \textsl{Density-Based Clustering Based on Hierarchical Density Estimates}. Proceedings of the 17th Pacific-Asia Conference on Knowledge Discovery in Databases (PAKDD 2013). Lecture Notes in Computer Science. \textbf{7819}, 160--172.
		\bibitem{}
		Chattopadhyay, S. and Maitra, R. (2017). \textsl{Gaussian-mixture-model-based cluster analysis finds five kinds of gamma-ray bursts in the BATSE catalogue}. Monthly Notices of the Royal Astronomical Society. \textbf{469}, 3374--3389.
		\bibitem{}
		Chattopadhyay, S. and Maitra, R. (2018). \textsl{Multivariate 
			$t$-mixture-model-based cluster analysis of BATSE catalogue
			establishes importance of all observed parameters, confirms five distinct
			ellipsoidal sub-populations of gamma-ray bursts}. Monthly Notices of the Royal Astronomical Society. \textbf{481}, 3196--3209.
		\bibitem{}
		Chattopadhyay, T., Misra, R., Chattopadhyay, A. K., Naskar, M. (2007). \textsl{Statistical evidence for three classes of gamma-ray bursts}. The Astrophysical Journal. \textbf{667}, 1017--1023.
		\bibitem{}
		Chen, X., Liu, W., Qiu, H., Lai J. (2011). \textsl{APSCAN: A parameter free algorithm for clustering}. Pattern Recognition Letters, \textbf{32}, 973--986.
		\bibitem{}
		Dezalay J.-P., Barat C., Talon R., Syunyaev R., Terekhov O., Kuznetsov A.:
		(1992), in Paciesas W. S., Fishman G. J., eds, AIP Conf. Ser. Vol. 265,
		Huntsville GRB Workshop. Am. Inst. Phys., New York, Page: 304.
		\bibitem{}
		Duda, R. O., Hart, P. E. and Stork, D. G. (2001). \textsl{Pattern Classification}, Wiley, New York.
		\bibitem{}
		Ester, M., Kriegel, H.-P., Sander, J. \& Xu, X. (1996). \textsl{A density-based algorithm for discovering clusters
			in large spatial databases with noise.} Proceedings of the Second International Conference on
		Knowledge Discovery and Data Mining (KDD-96). AAAI Press, Portland, Oregon, 226--231.
		\bibitem{}
		Everitt, B. S., Landau, S. and Leese,  M. (2001). \textsl{Cluster Analysis.} Arnold, London.
		\bibitem{}
		Frayley, C. and Raftery, A. E. (1998). \textsl{How Many Clusters? Which Clustering
			Method? Answers via Model-Based Cluster Analysis}. The Computer Journal. \textbf{41}, 578--588.
				\bibitem{}
			Gehrels, N., Ramirez-Ruiz, E., \& Fox, D. B. (2009). 
			\textsl{Gamma-Ray Bursts in the Swift Era.} Annual Review of Astronomy and Astrophysics. \textbf{47}, 567--617.
			\bibitem{}
			Ghosh, P. (2025). \textsl{A comparati v e study of Euclidean and spherical clustering techniques for BATSE gamma-ray burst data}.  Monthly Notices of the Royal Astronomical Society, \textbf{541}, 446–462.
				\bibitem{}
			Goldstein, A., Veres, P., Burns, E., Briggs, M. S., Hamburg,  R., Kocevski, D., Wilson-Hodge, C. A., Preece, R. D., Poolakkil, S.,  Roberts, O. J., Hui, C. M., Connaughton, V., Racusin, J., von Kienlin,  A., Canton, T. D., Christensen, N., Littenberg, T., Siellez,  K., Blackburn, L., Broida, J., Bissaldi, E., Cleveland, W. H., Gibby, M. H., Giles, M. M., Kippen, R. M. , McBreen, S., McEnery, J., Meegan, C. A., Paciesas, W. S. and Stanbro, M. (2017). \textsl{An Ordinary Short Gamma-Ray Burst with Extraordinary Implications: Fermi-GBM Detection of GRB 170817A}. The Astrophysical Journal Letters, \textbf{848}, Article id: L14, 14 Pages.
		\bibitem{}
		Gheyas, I., Parkinson, S. and Khan, S. (2021). \textsl{OCEAN: A Non-Conventional Parameter Free Clustering Algorithm Using Relative Densities of Categories}. International Journal of Pattern Recognition and Artificial Intelligence, \textbf{35}, 2150017.
		\bibitem{}
		Hakkila, J., Giblin, T. W., Roiger, R. J., Haglin, D. J., Paciesas, W. S., Meegan,
		C. A. (2003). \textsl{How Sample Completeness Affects Gamma-Ray Burst Classification.} The Astrophysical Journal. \textbf{582}, 320--329.
		\bibitem{}
		Hakkila, J., Haglin, D. J., Pendleton, G. N., Mallozzi, R. S., Meegan, C. A.,
		Roiger, R. J. (2000). \textsl{Gamma-ray burst class properties}. The Astrophysical Journal. \textbf{538}, 165--180.
		\bibitem{}
		Handl, J., Knowles, K. \& Kell, D. (2005). \textsl{Computational cluster validation in post-genomic data analysis}. Bioinformatics. \textbf{21}, 3201--3212.
		\bibitem{}
		Hahsler, M., Piekenbrock, M. and Doran, D. (2019). \textsl{dbscan: Fast density-based clustering with R}. Journal of
		Statistical Software. \textbf{91}, 1--30.
		\bibitem{}
		Hartigan, J. A. (1975). \textsl{Clustering Algorithms}. John Wiley \& Sons, New York, USA.
		\bibitem{}
		Hartigan, J. A. and Wong, M. A. (1979). \textsl{A K-means clustering algorithm}.
		Applied Statistics. \textbf{28}, 100--108.
		\bibitem{}
		Hastie, T., Tibshirani, R. and Friedman, J. (2001). \textsl{The Elements of Statistical Learning:
			Data Mining, Inference and Prediction}, Springer, New York.
		\bibitem{}
		Herrero, J., Valencia, A, and Dopazo, J. (2005). \textsl{A hierarchical unsupervised growing neural network for clustering gene expression patterns}. Bioinformatics, \textbf{17}, 126-136.
			\bibitem{}
		Horv\'{a}th, I. (1998). \textsl{A third class of gamma-ray bursts$?$}. The Astrophysical Journal. \textbf{508}, 757--759.
		\bibitem{}
		Horv\'{a}th, I. (2009). \textsl{Classification of BeppoSAX’s gamma-ray bursts}. Astrophysics \& Space Science. \textbf{323}, 83--86.
		\bibitem{}
		Horv\'{a}th, I., Bal\'{a}zs, L. G., Bagoly, Z., Ryde, F., M\'{e}sz\'{a}ros, A. (2006). \textsl{A new definition of the intermediate group of gamma-ray bursts.} Astronomy \& Astrophysics. \textbf{447}, 23--30.
		\bibitem{}
		Horv\'{a}th, I. and T\'{o}th, B. G. (2016). \textsl{The duration distribution of Swift Gamma-Ray Bursts.} Astrophysics \& Space Science. \textbf{361}, 155. 
		\bibitem{} 
		Horv\'{a}th, I., T\'{o}th, B. G., Hakkila, J., T\'{o}th, L. V., Bal\'{a}zs, L. G., R\'{a}cz, I. I., Pint\'{e}r,
		S. \& Bagoly, Z., (2018). \textsl{Classifying GRB 170817A/GW170817 in a Fermi duration–hardness
			plane}. Astrophysics \& Space Science, \textbf{363}, 53.
		\bibitem{}
		Huang, P. and Zeng, X. (2025). \textsl{Parameter-free Clustering with Adaptive Nearest Neighbors Learning}. Journal of Physics: Conference Series, \textbf{3004}, id: 012031.
		\bibitem{}
		Hubert, L. and Arabie, P. (1985). \textsl{Comparing partitions}. Journal of Classification, \textbf{2}, 193–218.
		\bibitem{}
		Jin, Z. P., Covino, S., Liao, N. H., Li, X., D’Avanzo, P., Fan, Y.-Z.,  Wei, D.-M. (2020). \textsl{A kilonova associated with GRB 070809}. Nature Astronomy. \textbf{4}, 77–82.
		\bibitem{}
		Johnson, R. A. and Wichern, D. W. (2007). \textsl{Applied Multivariate 
			Statistical Analysis}, Pearson Prentice Hall, New Jersey.
		\bibitem{}
		Kaufman, L. and Rousseeuw, P. J. (2005). \textsl{Finding Groups in Data: An Introduction to Cluster Analysis.} John Wiley and Sons, New Jersey.
		\bibitem{}
		King, A., Olsson, E., and Davies, M. B. (2007). \textsl{A new type of long gamma-ray burst.} Monthly Notices of the Royal Astronomical Society. \textbf{374}, L34.
		\bibitem{}
		Koonsanit, K., Jaruskulchai, C. and Eiumnoh, A. (2012). \textsl{Parameter-Free K-Means Clustering Algorithm for Satellite Imagery Application}. International Conference on Information Science and Applications: Suwon, Korea (South), pp. 1-6.
		\bibitem{}
		Kouveliotou, C., Meegan, C. A., Fishman, G. J., Bhat, N. P., Briggs, M. S., Koshut, T. M., Paciesas, W. S., \& Pendleton, G. N. (1993). \textsl{Identification of two classes of gamma-ray bursts.} The Astrophysical Journal. \textbf{413}, L101.
			\bibitem{}
		Kulkarni, S., and Desai, S. (2017). \textsl{Classification of gamma-ray burst durations using robust model-comparison techniques}.  Astrophysics and Space Science, \textbf{362}, Article no. 70.
		\bibitem{}
		Lamb, G. P., Tanvir, N. R., Levan, A. J., de Ugarte Postigo,  A., Kawaguchi, K., Corsi, A., Evans, P. A., Gompertz, B., Malesani, D. B., Page, K. L., Wiersema, K., Rosswog, S., Shibata, M., Tanaka, M., van der Horst, A. J., Cano, Z., Fynbo, J. P. U., Fruchter, A. S., Greiner, J., Heintz, K. E.,  Higgins, A., Hjorth, J., Izzo, L., Jakobsson, P., Kann, D. A., O'Brien, P. T., Perley, D. A., Pian, E., Pugliese, G., Starling, R. L. C., Thöne, C. C. , Watson, D., Wijers, R. A. M. J., and Xu, D. (2019). \textsl{Short GRB 160821B: A Reverse Shock, a Refreshed Shock, and a Well-sampled Kilonova}. The Astrophysical Journal, \textbf{883}, Article Number: 48, Pages: 12.
			\bibitem{}
		Levan, A., Crowther, P., de Grijs, R., Langer, N., Xu, D., Yoon, S.--C. (2016). \textsl{Gamma-Ray Burst Progenitors.}
		Space Science Reviews. \textbf{202}, 33--78.
		\bibitem{}
		Levan, A. J., Gompertz, B. P., Salafia, O. S. et al. (2024). \textsl{Heavy-element production in a compact object merger observed by JWST}. Nature, \textbf{626}, 737-741.
		\bibitem{}
		Lv, X., Ma, Y., He, X., Huang, H., Yang, J. (2018). \textsl{CciMST: A Clustering Algorithm Based on Minimum Spanning Tree and Cluster Centers}, Mathematical Problems in Engineering, Article ID 8451796, 14 pages
		\bibitem{}
			Luo, J.-W., Wang, F.-F., Zhu-Ge, J.-M., Li, Y., Zou, Y.-C., Zhang, B. (2023). \textsl{Identifying the Physical Origin of Gamma-Ray Bursts with Supervised Machine Learning}, The Astrophysical Journal, \textbf{959},
		Article No. 44, Pages: 17.
		\bibitem{}
		Marozzi, M. (2015). \textsl{Multivariate multidistance tests for high-dimensional low sample size case-control
			studies}. Statistics in Medicine, \textbf{34}, 1511–1526.
		\bibitem{}
		Matioli, L. C., Santos,  S. R., Kleina,  M., Leite, E. A. (2018). \textsl{A new algorithm for clustering based on kernel density estimation}. Journal of Applied Statistics. \textbf{45}, 347--366.
			\bibitem{}
		Mazets, E. P., Golenetskii, S. V., Ilyinskii, V. N., Panov, V. N., Aptekar, R. L., Guryan, Yu. A., Proskura, M. P., Sokolov, I. A., Sokolova, Z. Ya., Kharitonova, T. V., Dyatchkov, A. V., \& Khavenson, N. G (1981). \textsl{Catalog of cosmic gamma-ray bursts from the KONUS experiment data.} Astrophysics and Space Science. \textbf{80}, 119--143.
		\bibitem{}
		McLachlan, G. and Peel, D. (2000). \textsl{Finite Mixture Models}. John Wiley and Sons, New York.
			\bibitem{}
		Mehta, N. and Iyyani, S. (2024). \textsl{Exploring Gamma-Ray Burst Diversity: Clustering Analysis of the Emission Characteristics of Fermi- and BATSE-detected Gamma-Ray Bursts}. The Astrophysical Journal, \textbf{969}, Article id: 88, 12 pages.
			\bibitem{}
		Melandri, A., Malesani, D. B., Izzo, L., Japelj, J., Vergani, S. D., Schady, P., Sagués Carracedo, A., de Ugarte Postigo, A., Anderson, J. P., Barbarino, C., Bolmer, J., Breeveld, A., Calissendorff, P., Campana, S., Cano, Z., Carini, R., Covino, S., D’Avanzo, P., D’Elia, V., della Valle, M., De Pasquale, M., Fynbo, J. P. U., Gromadzki, M., Hammer, F., Hartmann, D. H., Heintz, K. E., Inserra, C., Jakobsson, P., Kann, D. A., Kotilainen, J., Maguire, K., Masetti, N., Nicholl, M., Olivares E., F., Pugliese, G., Rossi, A., Salvaterra, R., Sollerman, J., Stone, M. B., Tagliaferri, G., Tomasella, L., Thöne, C. C., Xu, D., Young, D. R. (2019). \textsl{GRB 171010A/SN 2017htp: a GRB-SN at z = 0.33}, Monthly Notices of the Royal Astronomical Society, \textbf{490}, 5366–5374.
		\bibitem{}
		Minaev, P. Y. and Pozanenko, A. S. (2020). \textsl{The $E_{p,i}–E_{iso}$ correlation: type I gamma-ray bursts and the new
			classification method}. Monthly Notices of the Royal Astronomical Society, \textbf{492}, 1919–1936.
		\bibitem{}
		Modak, S. (2019). \textsl{Uncovering astrophysical phenomena related to galaxies and other objects through statistical analysis.} Doctoral Thesis, University of Calcutta, URL: http://hdl.handle.net/10603/314773 
		\bibitem{}
		Modak, S. (2021). \textsl{Distinction of groups of gamma-ray bursts in the BATSE catalog through fuzzy clustering}. Astronomy and Computing. \textbf{34}, Article id 100441, 1--7.
		\bibitem{}
		Modak, S. (2022). \textsl{A new nonparametric interpoint distance-based measure for assessment of clustering}. Journal of Statistical Computation and Simulation. \textbf{9},	1062--1077.
		\bibitem{}
		Modak, S. (2023a). \textsl{Pointwise norm-based clustering of data in arbitrary dimensional space}. Communications in Statistics - Case Studies, Data Analysis and Applications, \textbf{9}, 121–134.
		\bibitem{}
		Modak, S. (2023b). \textsl{A new measure for assessment of clustering based on kernel	density estimation}. Communications in Statistics -- Theory and Methods, \textbf{52}, 5942-5951.
		\bibitem{}
		Modak, S. (2023c). \textsl{Validity index for clustered data in non-negative space}, Calcutta Statistical Association Bulletin, \textbf{75}, 60–71. 
		\bibitem{}
		Modak, S. (2024a). Book Review on \textsl{Finding Groups in Data: An Introduction to Cluster Analysis}. Journal of Applied Statistics, \textbf{51}, 1618--1620.
		\bibitem{}
		Modak, S. (2024b). \textsl{Evaluation of the number of clusters in a data set using p-values from multiple tests of hypotheses}. Communications in Statistics - Theory and Methods, \textbf{53}, 8878-8889.
		\bibitem{}
		Modak, S. (2024c). \textsl{A new interpoint distance-based clustering algorithm using kernel density estimation}, Communications in Statistics - Simulation and Computation, \textbf{53}, 5323-5341.
		\bibitem{} 
		Modak, S. (2024d), \textsl{Determination of the number of clusters through logistic regression analysis}, Journal of Applied Statistics, \textbf{51}, 2344-2363.
		\bibitem{}
		Modak, S. (2024e). \textsl{A New Clustering Accuracy Measure Based on Relative Distances and its Cross-Validation Using Dirichlet Distribution}, Journal of Statistical Theory and Practice, \textbf{18}, Article number 43.
		\bibitem{} 
		Modak, S. (2025), \textsl{Confirmation of binary clustering in gamma-ray bursts through an integrated p-value from multiple nonparametric tests of hypotheses}, Astronomy and Computing, \textbf{51}, Article id: 100931, 8 pages.
		\bibitem{}
		Modak, S. \& Bandyopadhyay, U. (2019). \textsl{A new nonparametric test for two sample multivariate location problem with application to astronomy}. Journal of Statistical Theory and Applications. \textbf{18}, 136--146.
		\bibitem{}
		Modak, S., Chattopadhyay, T., and Chattopadhyay, A. K. (2017).
				\textsl{Two phase formation of massive elliptical galaxies: study through cross-correlation including spatial effect}, Astrophysics and Space Science. \textbf{362}, Article id: 206, pages 1--10.
		\bibitem{}
		Modak, S., Chattopadhyay, A. K., and Chattopadhyay, T. (2018). \textsl{Clustering of gamma-ray bursts through kernel principal component analysis}. Communications in Statistics -- Simulation and Computation. \textbf{47}, 1088--1102.
		\bibitem{}
		Modak, S., Chattopadhyay, T., and Chattopadhyay, A. K. (2020). \textsl{Unsupervised classification of eclipsing binary light curves through k-medoids
			clustering}. Journal of Applied Statistics. \textbf{47}, 376--392.
		\bibitem{}
		Modak, S., Chattopadhyay, T. \& Chattopadhyay, A. K. (2022). \textsl{Clustering of eclipsing binary light curves through functional principal component analysis}. Astrophysics and Space Science. \textbf{ 367}, Article id: 19, pages 1--10.
		\bibitem{}
		Mu, B., Dai, M. and Yuan, S. (2020). \textsl{DBSCAN-KNN-GA: a multi Density-Level Parameter-Free clustering algorithm}. IOP Conference Series: Materials Science and Engineering, \textbf{715}, pp.~012023.
		 \bibitem{}
		Mukherjee, S., Feigelson, E. D., Babu, G. J., Murtagh, F., Fraley, C. \& Raftery, A.
		(1998). \textsl{Three types of gamma-ray bursts}. The Astrophysical Journal. \textbf{508}, 314--327.
		\bibitem{}
		Nakar, E. (2007). \textsl{Short-hard gamma-ray bursts}. Physics Reports. \textbf{442}, 166--236.
		\bibitem{}
		Norris, J. P., Cline, T. L., Desai, U. D., \& Teegarden, B. J. (1984). \textsl{Frequency of fast, narrow $\gamma$-ray bursts.} Nature. \textbf{308},  434--435.
		\bibitem{}
		Ng, A., Jordan, M., Weiss, Y. (2001). \textsl{On spectral clustering: analysis and an algorithm}. In: Dietterich,
		T., Becker, S., Ghahramani, Z. (eds.) Advances in Neural Information Processing Systems, 14,
		849--856. MIT Press, Cambridge.
		\bibitem{}
		Paczy\'{n}ski, B., (1986). \textsl{Gamma-ray bursters at cosmological distances
		}. The Astrophysical Journal. \textbf{308}, L43--L46.
		\bibitem{}
		Paczy\'{n}ski, B., (1998). \textsl{Are gamma-ray bursts in star-forming regions$?$}.  The Astrophysical Journal. \textbf{494}, L45.
		\bibitem{}
		Pakhiraa, M. K., Bandyopadhyay, S. and Maulik, U. (2004). \textsl{Validity index for crisp and fuzzy clusters}. Pattern Recognition. \textbf{37}, 487--501.
		\bibitem{}
		Qin, Y., Liang, E.-W., Liang, Y.-F., Yi, S.-X., Lin, L., Zhang, B.-B., Zhang, J., Lü, H.-J., Lu, R.-J., Lü, L.-Z., Zhang, B. (2013). \textsl{A Comprehensive Analysis of Fermi Gamma-Ray Burst Data. III. Energy-dependent $T_{90}$ Distributions of GBM GRBs and Instrumental Selection Effect on Duration Classification}. The Astrophysical Journal, \textbf{763}, Article id 15, Pages 9.
		\bibitem{}
		\v{R}\'{\i}pa, J. \& M\'{e}sz\'{a}ros, A. (2016). On the connection of gamma-ray bursts and X-ray flashes in the BATSE and RHESSI databases. Astrophysics \& Space Science. \textbf{361}, 370.
		\bibitem{} 
		Rossi, A., Rothberg, B., Palazzi, E., Kann, D. A., D'Avanzo, P., Amati, L., Klose, S., Perego, A., Pian, E., Guidorzi, C., Pozanenko, A. S., Savaglio, S., Stratta, G., Agapito, G., Covino, S., Cusano, F., D'Elia, V., De Pasquale, M., Della Valle, M., Kuhn, O., Izzo, L., Loffredo, E., Masetti, N., Melandri, A., Minaev, P. Y., Guelbenzu, A. N., Paris, D., Paiano, S, Plantet, C., Rossi, F., Salvaterra, R., Schulze, S., Veillet, C., Volnova, A. A. (2022). \textsl{The Peculiar Short-duration GRB 200826A and Its Supernova}. The Astrophysical Journal, \textbf{932}, Number 1.
		\bibitem{}
		Jun Yang, Shunke Ai, Bin-Bin Zhang, Bing Zhang, Zi-Ke Liu, Xiangyu Ivy Wang, Yu-Han Yang, Yi-Han Yin, Ye Li, Hou-Jun Lü (2022). \textsl{A long-duration gamma-ray burst with a peculiar origin}. Nature, \textbf{612}, 232–235.
		\bibitem{}
		Yang, Y. H., Troja, E., O’Connor, B., Fryer, C. L., Im, M., Durbak, J., Paek, G. S. H., Ricci, R., Bom, C. R., Gillanders, J. H., Castro-Tirado, A. J., Peng, Z.-K., Dichiara, S., Ryan, G., Eerten, van H., Dai, Z.-G., Chang, S.-W., Choi,  H., De, K., Hu, Y., Kilpatrick, C. D., Kutyrev, A., Jeong, M., Lee, C.-U., Pérez-García, I. (2024). \textsl{A lanthanide-rich kilonova in the aftermath of a long gamma-ray burst}. Nature, \textbf{626}, 742–745.
		\bibitem{}
		Rahman, M. A., Ang, L. M. and Seng, K. P. (2018) \textsl{Unique neighborhood set parameter 
			independent density-based clustering with outlier detection}. IEEE Access. \text{6}, 44707-44717.
			\bibitem{}
			Rajaniemi, H. J. \& M\"{a}h\"{o}nen, P. (2002). \textsl{Classifying Gamma-Ray Bursts using Self-organizing Maps.} The Astrophysical Journal. \textbf{566}, 202--209.
		\bibitem{}
			\v{R}\'{\i}pa, J.,  M\'{e}sz\'{a}ros, A., Veres, P., \& Park, I. H. (2012). \textsl{On the spectral lags and peak countsof the gamma-ray bursts detected by the RHESSI satellite}. The Astrophysical Journal, \textbf{756}, Article No. 44, 13 pages.
		\bibitem{}
		Ripley B. D. (1996). \textsl{Pattern recognition and neural networks}. Cambridge University Press, Cambridge.
		\bibitem{}
		Rousseeuw, P. J. (1987). \textsl{Silhouettes: A graphical aid to the interpretation and validation of cluster analysis.} Journal of Computational and Applied Mathematics. \textbf{20}, 53--65.
		\bibitem{}
		Roux, M. (2018). \textsl{A Comparative Study of Divisive and Agglomerative Hierarchical Clustering Algorithms}.
		Journal of Classification. \textbf{35}, 345–366. 
		\bibitem{}	
		Ruspini, E. H. (1970). \textsl{Numerical methods for fuzzy clustering.} Information Sciences. \textbf{2}, 319--350.
		\bibitem{}
		Sabarish B. A., Arunkumar C., Abineha P., Suruthi Lavanya, Deepak Menan (2025). \textsl{Efficient Fuzzy Trajectory Clustering Algorithm (EFTCA) for Electing Optimum Clusters}. Procedia Computer Science.
		\textbf{260}, 665-674.
		\bibitem{}
		Schölkopf, B., Smola, A. J. (2002). \textsl{Learning with Kernels: Support Vector Machines, Regularization, Optimization, and Beyond.} Cambridge: MIT Press.
		\bibitem{}
		Scrucca L., Fop M., Murphy T.B., Raftery A.E. (2016). \textsl{mclust 5: Clustering, Classification and Density Estimation Using Gaussian Finite Mixture Models}. The R Journal. \textbf{8}, 289-317.
		\bibitem{}
		Shahid, N. (2023). \textsl{Comparison of hierarchical clustering and neural network 
			clustering: an analysis on precision dominance}. Scientific Reports, \textbf{13}, Article number: 5661.
		\bibitem{}
		Silva, L. E. Brito Da, Melton, N. M. and Wunsch, D. C. (2020). \textsl{Incremental Cluster Validity Indices for Online Learning of Hard Partitions: Extensions and Comparative Study}. Institute of Electrical and Electronics Engineers, \textbf{8}, 22025--22047.
		\bibitem{}
		Steinley, D. (2004). \textsl{Properties of the Hubert-Arabie adjusted Rand index. Psychological Methods}, \textsl{9}, 386–396.
		\bibitem{}
		Silva, L. E. Brito Da, Melton, N. M. and Wunsch, D. C. (2020). \textsl{Incremental Cluster Validity Indices for Online Learning of Hard Partitions: Extensions and Comparative Study}. Institute of Electrical and Electronics Engineers, \textbf{8}, 22025--22047.
					\bibitem{}
				Tarnopolski, M. (2019). \textsl{Analysis of the Duration–Hardness Ratio Plane of Gamma-Ray Bursts Using Skewed
						Distributions}, The Astrophysical Journal. \textbf{870}, Article id: 105, 9 pages.
		\bibitem{}
		Tarnopolski, M. (2022). \textsl{Graph-based clustering of gamma-ray bursts}. Astronomy \& Astrophysics, \textbf{657}, Article No. A13, 8 pages.
		\bibitem{}
		Tóth, B. G., Rácz, I. I. and  Horváth, I. (2019). \textsl{Gaussian-Mixture-Model-based Cluster Analysis of
		Gamma-Ray Bursts in the BATSE Catalog.} Monthly Notices of the Royal Astronomical Society,
				\textbf{486}, 4823–4828.
				\bibitem{}
					Troja, E., Castro-Tirado, A. J., González, J. B., Hu, Y., Ryan, G. S., Cenko, S. B., Ricci, R., Novara, G., Sánchez-Rámirez, R., Acosta-Pulido, J. A., Ackley, K. D.,  García, M. D. C., Eikenberry, S. S., Guziy, S., Jeong, S., Lien, A. Y., Márquez,  I., Pandey, S. B., Park, I. H., Sakamoto, T., Tello, J. C., Sokolov, I. V., Sokolov,  V. V., Tiengo, A., Valeev, A. F., Zhang, B. B., Veilleux, S. (2019). \textsl{The afterglow and kilonova of the short GRB 160821B.} Monthly Notices of the Royal Astronomical Society, \textbf{489}, 2104–2116.
\bibitem{}
Troja, E., Fryer, C. L., O'Connor, B., Ryan, G., Dichiara, S., Kumar, A., Ito, N., Gupta, R., Wollaeger, R. T., Norris, J. P., Kawai, N., Butler, N. R., Aryan, A., Misra, K., Hosokawa, R., Murata, K. L., Niwano, M., Pandey, S. B., Kutyrev, A., van Eerten, H. J., Chase, E. A., Hu, Y.-D., Caballero-Garcia, M. D., Castro-Tirado, A. J. (2022). \textsl{A nearby long gamma-ray burst from a merger of compact objects}. Nature, \textbf{612}, 228-231.
 
						\bibitem{}
					Tsutsui, R. \& Shigeyama, T. (2014). \textsl{On the subclasses in Swift long gamma-ray bursts: A clue to different central engines.} Publications of the Astronomical Society of Japan. \textbf{66}, 42.
						\bibitem{}
					Usov, V. V. (1992). \textsl{Millisecond pulsars with extremely strong magnetic fields as a cosmological source of $\gamma$-ray bursts}. Nature. \textbf{357}, 472--474.
						\bibitem{}
					Wang, H., Zhang, F.-W., Wang,  Y.-Z., Shen, Z.-Q., Liang, Y.-F., Li,  X., Liao, N.-H., Jin, Z.-P., Yuan,  Q., Zou, Y.-C., Fan, Y.-Z., and Wei, D.-M. (2017), \textsl{The GW170817/GRB 170817A/AT 2017gfo Association: Some Implications for Physics and Astrophysics}. The Astrophysical Journal Letters, \textbf{851}, Article id: L18, 7 Pages.
\bibitem{}
Wei, S.-Q., Xiao, S., Zhang, Y.-Q., Jiang, Z.-H., Liao, T.-L., Wang,  M.-Z., Wen,  Y., Zhang, D.-Y., Zhang, S.-J., Li, X., Xiong, S.-W., Zhang, Y., You, Z.-Y., Xiao, W.-J. (2026). \textsl{Uncovering Anomalous Gamma-Ray Bursts beyond Duration-based Classification}. The Astrophysical Journal Letters, \textbf{1003}, Article id: L13, 7 pages.
						\bibitem{}
					Woosley, S. E. (1993). \textsl{Gamma-Ray Bursts from Stellar Mass Accretion Disks around Black Holes.} The Astrophysical Journal. \textbf{405}, 273--277.
					\bibitem{}
					Woosley, S. E. \& Bloom, J. S. (2006). \textsl{The Supernova Gamma-Ray Burst Connection}. Annual Review of Astronomy \& Astrophysics. \textbf{44}, 507--556.
		\bibitem{}
		Yang, E. B., Zhang, Z. B., \& Jiang, X. X. (2016). \textsl{Two dimensional classification of the Swift/BAT GRBs}.  Astrophysics and Space Science, \textbf{361}, Article id: 257.
		\bibitem{}
		Yang, J., Ai, S., Zhang, B.-B., Zhang, B., Liu, Z.-K., Wang,  X. I., Yang,  Y.-H., Yin, Y.-H., Li, Y., Lü, H.-J. (2022). \textsl{A long-duration gamma-ray burst with a peculiar origin}. Nature. \textbf{612}, 232–235.
		\bibitem{}
		Yata, K., and Aoshima, M. (2010). \textsl{Effective PCA for high-dimension, low-sample-size data with singular value decomposition of cross data matrix}. Journal of Multivariate Analysis, \textbf{101}, 2060–2077.
		\bibitem{}
		Zhang, B., Zhang, B.-B., Virgili, F. J., Liang, E.-W., Kann, D. A., Wu, X.-F., Proga, D., Lv, H.-J., Toma, K., Mészáros, P., Burrows, D. N., Roming, P. W. A., Gehrels, N. (2009). \textsl{Discerning the Physical Origins of Cosmological Gamma-ray Bursts Based on Multiple Observational Criteria: The Cases of z = 6.7 GRB 080913, z = 8.2 GRB 090423, and Some Short/Hard GRBs}. The Astrophysical Journal, \textbf{703}, Number 2.
		\bibitem{}
		Zhang, F.--W., Shao, L., Yan J.--Z., \& Wei, D.--M. (2012). \textsl{Revisiting the Long/Soft-Short/Hard Classification of Gamma-Ray Bursts in the Fermi Era.} The Astrophysical Journal, \textbf{750}, 88.
	\bibitem{}
	Zhu, J.-P., Wang, X. I., Sun, H., Yang, Y.-P., Li, Z., Hu, R.-C., Qin, Y., Wu, S. (2022). \textsl{Long-duration Gamma-Ray Burst and Associated Kilonova Emission from Fast-spinning Black Hole-Neutron Star Mergers}. The Astrophysical Journal Letters, \textbf{936}, Article id L10.		
			\bibitem{}
			Zhu, S.-Y., Sun, W.-P., Ma, D.-L. and Zhang, F.-W. (2024). \textsl{Classification of Fermi gamma-ray bursts based on machine learning.}Monthly Notices of the Royal Astronomical Society. \textbf{532}, 1434–1443.
			\bibitem{}
			Zhu, S.-Y., Shao, L., Tam, P.-H. T., Zhang, F.-W. (2025). \textsl{Unsupervised machine learning classification of gamma-ray bursts
				based on the rest-frame prompt emission parameters}. Astronomy \& Astrophysics, \textbf{702}, A173.
			\bibitem{}
			Zitouni, H., Guessoum, N., Azzam, W. J., \& Mochkovitch, R. (2015). \textsl{Statistical study of observed and intrinsic durations among
				BATSE and Swift/BAT GRBs.} Astrophysics and Space Science. \textbf{357}, 7.
		
	\end{thebibliography}
\end{document}